\documentclass{article}
\usepackage{ijcai21}

\usepackage{times}
\usepackage{soul}
\usepackage{url}
\usepackage[hidelinks]{hyperref}
\usepackage[utf8]{inputenc}
\usepackage[small]{caption}
\usepackage{graphicx}
\usepackage{booktabs}       
\usepackage{amsfonts}       
\usepackage{xcolor}

\usepackage[linesnumbered,titlenumbered,ruled,vlined,commentsnumbered,noend]{algorithm2e}
\usepackage{algpseudocode}
\usepackage[tight]{subfigure}

\usepackage{amssymb}
\usepackage{amsmath}
\usepackage{amsthm}
\usepackage{multirow}
\usepackage[switch]{lineno}  %

\newtheorem{definition}{Definition}
\newcommand{\AlgName}{\textsc{PST}\xspace}

\newcommand{\bheading}[1]{{\vspace{2pt}\noindent{\textbf{#1}}\hspace{2pt}}}

\newenvironment{packeditemize}{
\begin{list}{$\bullet$}{
\setlength{\labelwidth}{8pt}
\setlength{\itemsep}{0pt}
\setlength{\leftmargin}{\labelwidth}
\addtolength{\leftmargin}{\labelsep}
\setlength{\parindent}{0pt}
\setlength{\listparindent}{\parindent}
\setlength{\parsep}{0pt}
\setlength{\topsep}{3pt}}}{\end{list}}

\title{Fine-tuning Is Not Enough: \\A Simple yet Effective Watermark Removal Attack for DNN Models}

\author{
Shangwei Guo$^1$
\and
Tianwei Zhang$^2$\footnote{Tianwei Zhang is the corresponding author.}\and
Han Qiu$^3$\and
Yi Zeng$^4$\and
Tao Xiang$^1$ \And
Yang Liu$^2$
\affiliations
$^1$College of Computer Science, Chongqing University, China\\
$^2$School of Computer Science and Engineering, Nanyang Technological University, Singapore\\
$^3$Institute for Network Sciences and Cyberspace, Tsinghua University, China\\
$^4$The Bradley Department of Electronic and Computer Engineering, Virginia Tech Blacksburg, USA
\emails
\{swguo,txiang\}@cqu.edu.cn,
\{tianwei.zhang,yangliu\}@ntu.edu.sg,
qiuhan@tsinghua.edu.cn,
yizeng@vt.edu
}

\begin{document}

\maketitle

\begin{abstract}
    Watermarking has become the tendency in protecting the intellectual property of DNN models.
    Recent works, from the adversary's perspective, attempted to subvert watermarking mechanisms by designing watermark removal attacks. However, these attacks mainly adopted sophisticated fine-tuning techniques, which have certain fatal drawbacks or unrealistic assumptions.
    In this paper, we propose a novel watermark removal attack from a different perspective. Instead of just fine-tuning the watermarked models, we design a simple yet powerful transformation algorithm by combining imperceptible pattern embedding and spatial-level transformations, which can effectively and blindly destroy the memorization of watermarked models to the watermark samples. We also introduce a lightweight fine-tuning strategy to preserve the model performance. Our solution requires much less resource or knowledge about the watermarking scheme than prior works. Extensive experimental results indicate that our attack can bypass state-of-the-art watermarking solutions with very high success rates. Based on our attack, we propose watermark augmentation techniques to enhance the robustness of existing watermarks.

\end{abstract}

\section{Introduction}\label{sec:introduction}

Watermarking, originally designed for digital media \cite{petitcolas1999information},  has been recently applied to the protection of Deep Learning (DL) models. This technique enables the ownership verification of DL models by embedding watermarks into the Deep Neural Networks (DNNs). One promising approach is data-poisoning watermarking \cite{zhang2018protecting,adi2018turning,fan2019rethinking,zhang2020model}, which modifies the model to give pre-defined output on some carefully-crafted watermark samples. Then the owners with only black-box accesses to the suspicious model can extract such watermarks by querying the model.

Similar to digital watermarking, a satisfactory data-poisoning watermarking mechanism must satisfy two requirements. The first one is \emph{functionality-preserving}: the embedded watermarks should not affect the performance of the target model on normal samples. The second is \emph{robustness}, where the watermarks cannot be removed with common model transformation, e.g., fine-tuning, model compression, etc. Even if the adversary knows the target model is watermarked, he has no means to remove the watermarks if he does not know the details of watermark samples.

Whether existing watermarking solutions are practically robust is widely challenged. A quantity of works try to invalidate these watermarking mechanisms and can be classified into two categories. The first strategy is to \emph{detect} the verification queries and manipulate the responses. Some works \cite{namba2019robust,aiken2020neural} applied the strategies of backdoor attack detection to watermark detection, as some watermark designs are based on the backdoor techniques. However, these solutions fail to defeat backdoor attacks (as well as watermarks) with complex patterns \cite{liu2019abs,tan2019bypassing}. They are not applicable to watermark schemes based on other techniques (e.g., adversarial examples, out-of-distribution samples) either.

Another strategy is to directly \emph{remove} the watermarks via model transformation. Although most watermarking solutions claim to be robust against various model transformations (e.g., fine-tuning, compression), recent works \cite{chen2019refit,shafieinejad2019robustness,liu2020removing} attempted to break such statement by introducing advanced fine-tuning methods with data augmentation, sophisticated loss function, etc. Unfortunately, these works suffer from significant limitations. For instance, \cite{shafieinejad2019robustness} requires original training samples to fine-tune the watermarked model. \cite{chen2019refit} needs to know the type of the watermark to adjust the fine-tuning parameters. These assumptions make the attacks less unrealistic or practical.


It seems not promising to just use model fine-tuning for effective watermark invalidation. As a result, we propose a new watermark removal attack from a different direction: inference sample transformation. The key insight of our solution is that \emph{watermark samples are less robust than normal samples}. Hence, we can design a preprocessing function to compromise the verification output while not affecting the normal output. Specifically, our novel function \AlgName consists of a series of transformations (scaling, embedding random imperceptible patterns, spatial-level transformations).
We further fine-tune the watermarked model with unlabelled out-of-distribution data to maintain the model's performance.

We are the \textit{first} to explore the possibility of inference sample transformation for watermark removal. Our solution has much fewer requirements with more reasonable assumptions than prior works: it does not need any samples from the original training set (as in \cite{shafieinejad2019robustness}) or even following the original distribution (as in \cite{chen2019refit,liu2020removing}). It is unified and does not need to know the adopted watermark mechanism for hyperparameter adjustment (as in \cite{chen2019refit}).
Its effectiveness is also confirmed by extensive evaluations.

Given the severity and practicality of our attack, we further introduce a defense methodology to enhance the robustness of watermarks. Inspired by the data augmentation technique which can enhance the model performance and generalization, we propose to augment the watermark samples with various transformations during the embedding process. In this case, the verification results remain correct even the adversary performs different preprocessing over the inference samples. Evaluations indicate this strategy can significantly improve the robustness of existing watermarking mechanisms.

\section{Background and Related Works}\label{sec:problemdef}
\subsection{Watermarking DNN Models}

Existing watermarking schemes for DNNs can be classified into two categories. The first category is parameter-embedding watermarking \cite{uchida2017embedding,rouhani2019deepsigns}, which embed watermarks into the parameters without decreasing the model's performance. Verification of such watermarks require white-box access to the target models, which may not be achievable in some scenarios.

We are more interested in the second category, data-poisoning watermarking \cite{le2019adversarial,zhang2018protecting,adi2018turning}. These solutions take a set of carefully crafted sample-label pairs as watermarks and embed their correlation into DL models during the training process. We formally define the data-poisoning DNN watermarking scheme and illustrate the necessary properties.

\begin{definition}
    A DNN watermarking scheme with data poisoning is a tuple of probabilistic polynomial time algorithms (\textbf{WMGen}, \textbf{Mark}, \textbf{Verify}), where
    \begin{packeditemize}
        \item \textbf{WMGen} generates a set of watermarks $W = \{(x_i, y_i)\}_{i=1}^n$, in which $x_i$ is a secret watermark input and $y_i$ is the corresponding verification label.
        \item \textbf{Mark} embeds the watermarks into a DL model $f$ and outputs the watermarked model $\widehat{f}$ such that $\widehat{f}(x_i) = y_i$ for $\forall \ (x_i, y_i) \in W$.
        \item \textbf{Verify} sends $\{x_i\}_{i=1}^n$ to a DL model $\widetilde{f}$ and obtains the predictions $\{y_{\widetilde{f},i}\}_{i=1}^n$. If the probability that $y_{\widetilde{f},i}$ equals $y_i$ for $i \in [1, n]$ is larger than a predefined value $\tau$, \textbf{Verify} outputs 1. Otherwise it outputs 0.
    \end{packeditemize}
\end{definition}

Let $P_{\widehat{f}, D}$ be the prediction accuracy of $\widehat{f}$ on the normal samples following the  data distribution $D$; $P_{\widehat{f} , W}$ be the accuracy of $\widehat{f}$ on the verification watermarks $W$, i.e.,
\begin{align}
    P_{\widehat{f},D} &= Pr(\widehat{f}(x) == f(x), x \backsim D),\\
    P_{\widehat{f},W} &= Pr(\widehat{f}(x_i) == y_i, \{x_i, y_i\} \in W).
\end{align}

A data-poisoning watermarking scheme should have the following basic properties: 
(1) \emph{Functionality-preserving}: $P_{\widehat{f},D}$ should be identical with $P_{f,D}$. (2) \emph{Robustness}: \textit{\textbf{Verify}} outputs 1 even if $\widehat{f}$ is slightly transformed to a different model $\widetilde{f}$ using common techniques such as fine-tune or model compression, i.e. $P_{\widetilde{f},W} > \tau$.

Various state-of-the-art data-poisoning techniques have been proposed to embed watermarks into DNN models. These methods can be classified into three categories:

\begin{packeditemize}
    \item \bheading{Perturbation-based}: the samples are generated by slightly perturbing normal samples, where the difference between the two types of samples is bounded. They can be generated using the adversarial frontier stitching \cite{le2019adversarial} or GAN-based techniques \cite{li2019prove}.

    \item \bheading{Pattern-based}: the samples are generated by embedding a certain pattern into normal images using backdoor techniques \cite{adi2018turning,zhang2018protecting}.

    \item \bheading{Out-of-distribution (OOD)}: the samples are randomly selected from the internet and totally different from the normal ones \cite{adi2018turning}.

\end{packeditemize}

\subsection{Attacking Watermarking Schemes}

We consider a threat model where the adversary has white-box accesses to the target model and is able to alter the model. He aims to break the property of Equation (2) but still maintain the property of Equation (1).
A good removal attack must meet the following requirements.

\begin{packeditemize}
    \item \emph{Comprehensive}: the attack should be able to invalidate all existing state-of-the-arts watermarking solutions, including perturbation-based, pattern-based, and OOD.

    \item \emph{Watermark-agnostic}: the adversary has no knowledge about the employed watermarking scheme, including the technique and watermark type. He has no access to the original training samples, or the distribution.

    \item \emph{Efficient}: the adversary should be able to remove the watermarks in a lightweight manner, i.e., with much less computation cost than training a model from scratch. Otherwise, he will lose the motivation of stealing the target model, and just train his own copy.

\end{packeditemize}

Existing watermark attacks fail to satisfy all these requirements. These solutions can be classified into two categories. The first approach is to distinguish the watermark samples from normal samples during the inference process, and then alter the verification results \cite{namba2019robust,aiken2020neural}. However, these attacks are not watermark-agnostic, as they require the entire training dataset.
They are not comprehensive as they only work for simple pattern-based watermarks.

The second is to remove watermarks from the model using fine-tuning techniques \cite{chen2019refit,shafieinejad2019robustness,liu2020removing}. They carefully designed learning rate schedule and synthesized fine-tuning samples to make the watermarked model forget the watermark samples. Unfortunately, these approaches also suffers from several fatal limitations: they are not watermark-agnostic, as they need to access the original training data \cite{shafieinejad2019robustness,chen2019refit,liu2020removing}. \cite{chen2019refit} also needs to know the watermark type for hyperparameter adjustment. \cite{liu2020removing} is not comprehensive to cover the OOD watermarking scheme.

\begin{figure}[t]
	\centering
	\includegraphics[width=0.9\columnwidth]{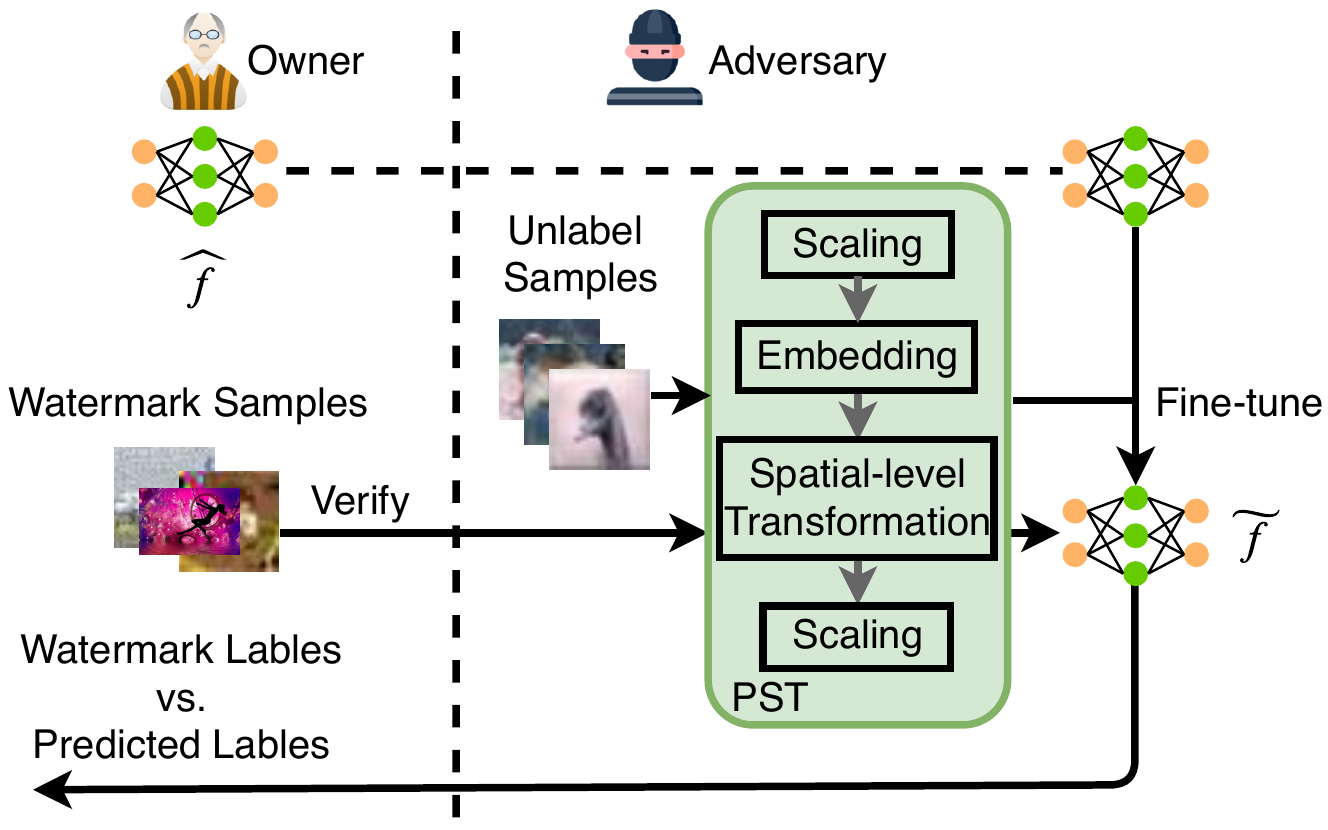}
	\caption{Overview of the proposed attack.}
	\label{fig:attack}
\end{figure}

\section{Proposed Attack}\label{sec:algorithm}
Figure \ref{fig:attack} illustrates the workflow of our proposed attack. The core of our solution is a powerful function, \textbf{P}attern embedding and \textbf{S}patial-level \textbf{T}ransformation (\AlgName). Given a watermarked model $\widehat{f}$, the adversary first preprocesses some OOD data samples with \AlgName, and uses them to fine-tune $\widehat{f}$ for a few epochs. Then he can serve the fine-tuned model $\widetilde{f}$ online. During inference, he applies \AlgName for each sample and feeds it to $\widetilde{f}$. The model output of a normal sample will remain the same while the watermark ones will be altered.

\noindent\textbf{Key Insight.} Our approach is based on the robustness gap between watermark and normal samples. Intuitively, a data-poisoning watermarked model is trained to memorize the relationships between $\{x_i\}_{i=1}^n$ and $\{y_i\}_{i=1}^n$. During the verification, the watermarked model will predict $y_i$ given $x_i$. However, such memory may be fragile because $y_i$ has to be different from the ground truth of $x_i$. Based on this, \AlgName is expected to preprocess watermark samples, and make them unrecognizable by $\widehat{f}$. To maintain the model's performance over normal samples, we leverage \AlgName to preprocess some OOD samples and fine-tune the watermarked model.

\subsection{\AlgName}

Our \AlgName function consists of three modules: (1) \emph{scaling} is to resize an input sample to a fixed size; (2) \emph{imperceptible pattern embedding} introduces human-unnoticeable perturbations to reduce the impact of abnormal pixels and patterns in the watermark samples; (3) \emph{spatial-level transformations} further destroy the special spatial shapes introduced by the watermarking scheme.
Detailed steps are illustrated below.

\bheading{Step 1: Scaling.}
Given a sample $x$ of an arbitrary size, \AlgName first scales it to a fixed size. Specifically, \AlgName chooses the bicubic interpolation \cite{meijering2002chronology} for scaling to reduce the correlation between watermark samples and labels. This operation $\texttt{Bicubic}$ interpolates pixels by considering a small square neighborhood (4$\times$4 in our experiments): $x = \texttt{Bicubic}(x, \beta),$ where $\beta$ is the scaling paramater. We choose this bicubic interpolation
because samples with bicubic interpolation have fewer interpolation artifacts and often outperform the ones with bilinear or nearest-neighbor interpolation. Thus, this step can better preserve the performance of $\widehat{f}$ on normal samples.

\bheading{Step 2: Imperceptible Pattern Embedding.}
Besides the bicubic interpolation, \AlgName embeds random imperceptible patterns into the scaled watermark samples to further affect the prediction of $\widehat{f}$. Due to the imperceptibility property, the embedded pattern will not impact the performance of $\widehat{f}$ over normal samples. Since the adversary has no prior knowledge about the watermarking scheme and watermark samples, we design a random yet easily identifiable pattern to alter the memory of $\widehat{f}$ about the watermark samples.

In particular, we adopt a random grid median filter \texttt{MedianFilter} to generate the desired imperceptible pattern. During the embedding, \AlgName first selects a set of random rows and columns with the same interval size $v$. For each pixel $x_{i,j}$ in the selected rows and columns, it is replaced by the pixel with the median value in the neighborhood. The pattern size is controlled by the interval size $v$: a large $v$ can lead to a small number of the selected columns and rows, and a more imperceptible pattern. This median filter can be replaced by other types of filters such as the maximum filter.

\bheading{Step 3: Spatial-level Transformations.}
Although the above pixel-level transformations can affect the watermark samples with small perturbations, they are insufficient to remove large perturbations such as pattern-based watermarks. To this end, we propose to adopt both linear and nonlinear spatial-level transformations to further compromise the effects of watermark samples.
We integrate two different transformations and use a parameter $\gamma$ to uniformly control the strength of each one. Specifically, Let $(i, j)$ be a pixel of $x$ and $(\widetilde{i}, \widetilde{j})$ be the corresponding transformed pixel. For each operation, we require the distance between each dimension of the input and output pixels to be lower than $\gamma$, i.e., $$|i -\widetilde{i}| \leq \gamma, \ \ \text{and} \ \ |j -\widetilde{j}| \leq \gamma.$$

Our \AlgName first adopts random affine transformations (\texttt{Affine}) over the pattern-embedded $x$. An affine transformation is linear and can be formalized as a $2\times 3$ matrix, which preserves points, straight lines, and planes of samples. For a pixel $(i, j)$, it calculates
\begin{equation}
	\begin{bmatrix}
	\widetilde{i}\\
	\widetilde{j}
	\end{bmatrix} = \begin{bmatrix}
	a_{11} & a_{12} & b_{1}\\
	a_{21} & a_{22} & b_{2}
	\end{bmatrix}\begin{bmatrix}i\\j\\1\end{bmatrix}.
\end{equation}

For example, the matrix of the translation transformation in the horizontal direction is $\begin{bmatrix}
	1 & 0  & \phi\\
	0 & 1 & 0
\end{bmatrix}$. To meet the distance constraint, we randomly choose $\phi$ such that $|\phi| \leq \gamma$ for all the columns. We can combine multiple affine transformations together such as rotation and translation.

Besides affine transformations, we also adopt an elastic transformation \texttt{Elastic} to further modify the samples, which is a nonlinear transformation that produces random elastic distortions. It first generates random displacement fields that are convolved with a Gaussian distribution of standard deviation $\sigma$. The displacement fields are then multiplied by a scaling factor $\alpha$ that controls the intensity of the deformation.
We adjust the standard deviation $\sigma$ and the scaling factor $\alpha$ to restrict the displacement to be smaller than $\gamma$.

\subsection{Attack Pipeline}
Algorithm \ref{algorithm} describes the details of the two-stage attack:

\bheading{Offline fine-tuning}(Lines \ref{line:fine-begain}-\ref{line:fine-end}).
The adversary first adopts \AlgName to fine-tune $\widehat{f}$ to enhance the model performance. Different from prior works \cite{chen2019refit,shafieinejad2019robustness,liu2020removing} which require the original labeled training data or distribution, our method can select an arbitrary set of unlabeled and out-of-distribution samples $\{x_i'\}$. The adversary first calculates the corresponding label of each sample using $\widehat{f}$ as an oracle. Then he establishes a dataset $\{(g^{pst}(x_i'), \widehat{f}(g^{pst}(x_i')))\}$ to fine-tune the model $\widehat{f}$ for a few epochs to reach the desired performance. This fine-tuning process takes very short time to complete, and can effectively adjust the noise and perturbations introduced by \AlgName.

\bheading{Online inference}(Lines \ref{line:extract-begain}-\ref{line:extract-end}).
For each sample $x$, the online model returns $\widetilde{y} = \widetilde{f}(g^{pst}(x))$. When the owner executes the \textbf{\textit{Verify}} function to query the model with a watermark sample $x_i$,  $\widetilde{y}_i$ is different from $y_i$ with a high probability, due to the effectiveness of $g^{pst}$ and $\widetilde{f}$. Thus, the similarity between $\{y_i\}_{i=1}^{n}$ and $\{\widetilde{y}_i\}_{i=1}^{n}$ would be smaller than $\tau$, which can prevent the model owner from identifying the watermarks. In contrast, $\widetilde{f}$ will give the correct results for normal samples $x$.

\begin{algorithm}[t!]
	\SetAlgoLined
	\SetKwProg{Fn}{Function}{:}{}

    \SetKwInOut{Input}{Parameters}
    \Input{$\beta$, $v$, $\gamma$, $\widehat{f}$, $\{x_i'\}$, $\{x_i, y_i\}_{i=1}^n$}

	\SetKwFunction{FMain}{\textbf{\textit{Verify}}}
    \Fn{\FMain{$\{x_i, y_i\}_{i=1}^n$}\label{line:extract-begain}}{
	\ForEach{$x_i \in \{x_i\}_{i=1}^{n}$}{
		$\widetilde{y}_i \gets \widetilde{f}(g^{pst}(x_i))$, where $\widetilde{f} \gets \texttt{Fine-tune}$\;
	}
	\If{$d(\{y_i\}_{i=1}^{n}, \{\widetilde{y}_i\}_{i=1}^{n}) < \tau$}{
		\Return 1
	}\Else{
		\Return 0 \label{line:extract-end}
	}
	}
	\BlankLine

	\SetKwFunction{Fpst}{$g^{pst}$}
	\Fn{\Fpst{$x$}\label{line:pst-begain}}{
        $x \gets \texttt{Bicubic}(x, \beta)$\;
		Randomly selected rows and columns with interval $v$\;
		\ForEach{$x_{i,j}$ \emph{in the selected rows and columns}}{
			$x_{i,j} \gets \texttt{MedianFilter}(x_{i,j})$\;
		}
		$x \gets \texttt{Affine}(x, \gamma)$\;
		$x \gets \texttt{Elastic}(x, \gamma)$\;
		$\widetilde{x} \gets \texttt{Bicubic}(x, \frac{1}{\beta})$\;
		\Return $\widetilde{x}$\label{line:pst-end}

	}
	\BlankLine
	\SetKwFunction{FMain}{Fine-tune}
    \Fn{\FMain{$\widehat{f}, \{x_i'\}$}\label{line:fine-begain}}{
        \ForEach{$x_i' \in \{x_i'\}$}{
			$y_i' \gets \widehat{f}(g^{pst}(x_i'))$\;
		}
		$\widetilde{f} \gets$ \texttt{Train}($\widehat{f}, \{g^{pst}(x_i'), y_i'\}$)\;
        \Return $\widetilde{f}$ \label{line:fine-end}
	}
	\caption{Our watermark removal attack.}\label{algorithm}
\end{algorithm}
\section{Experiments}\label{sec:experiments}
\subsection{Experimental Setup}
\bheading{Datasets and DNN models.} Our attack is model-agnostic and dataset-agnostic. Following existing works, we conduct experiments on the CIFAR datasets with the same number of training samples.
We evaluate our attack on the ResNet model \cite{he2016deep}. We adopt a batch size of 100 and the learning rate decay with an initialize value of 0.2.

\bheading{Watermark samples.} We evaluate our attack on four types of state-of-the-art watermark samples: 1) Perturbation: the watermark samples are generated by adding imperceptible perturbations on normal samples. They are close to the frontiers of the models \cite{le2019adversarial}. 2)  Patch: the watermark samples are created by embedding a specific patch into the normal samples \cite{zhang2018protecting}. 3) Content: the watermark samples are generated by embedding the word ``TEXT'' into the normal samples \cite{zhang2018protecting}. 4) OOD: the watermark samples are randomly selected from unrelated datasets \cite{adi2018turning}.

\begin{figure}[t]
    \centering
    \includegraphics[width=0.19\columnwidth]{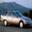}
    \includegraphics[width=0.19\columnwidth]{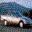}
    \includegraphics[width=0.19\columnwidth]{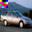}
    \includegraphics[width=0.19\columnwidth]{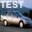}
	\resizebox{\width}{0.19\columnwidth}{\includegraphics[width=0.19\columnwidth]{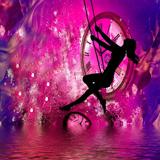}}
    \includegraphics[width=0.19\columnwidth]{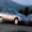}
    \includegraphics[width=0.19\columnwidth]{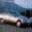}
    \includegraphics[width=0.19\columnwidth]{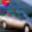}
    \includegraphics[width=0.19\columnwidth]{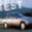}
    \resizebox{\width}{0.19\columnwidth}{\includegraphics[width=0.19\columnwidth]{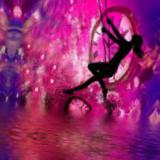}}
    \includegraphics[width=0.19\columnwidth]{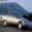}
    \includegraphics[width=0.19\columnwidth]{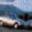}
    \includegraphics[width=0.19\columnwidth]{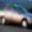}
    \includegraphics[width=0.19\columnwidth]{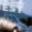}
    \resizebox{\width}{0.19\columnwidth}{\includegraphics[width=0.19\columnwidth]{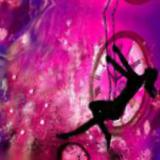}}
	\caption{Examples of the watermark samples in our experiments. First row: normal sample and the four types of watermark samples; Second row: the samples preprocessed using \AlgName with $\gamma =10$; Third row: the samples preprocessed using \AlgName with $\gamma =20$.}
	\label{fig:watermarks}
\end{figure}

Figure \ref{fig:watermarks} (first row) illustrates an example of a normal sample and the corresponding four types of watermark samples. Following the implementation in \cite{adi2018turning,le2019adversarial}, we set the number of watermark samples as 100.  We set $\tau$ as 60\% for both CIFAR10 and CIFAR100 when perturbation-based watermark samples are involved. Otherwise, $\tau$ is 20\% for CIFAR10 and 10\% for CIFAR100.
We train models from scratch to embed the watermarks. Each training process is stopped after 60 epochs if the model accuracy on watermark samples is 1. Otherwise, we continue the training till all watermarks are embedded.

\bheading{Implementation Details.}
We \emph{uniformly} set the parameters for all datasets and watermarks, which indicates our attack does not require any prior knowledge of the watermarking mechanisms.
In particular, for \AlgName, we set $\beta$ as 5 and the size of the scaled images is 160$\times$160. We set the interval size as 5 to generate the imperceptible pattern. We set the displacement bound $\gamma$ in spatial-level transformations as 15.
We use both in-distribution and out-of-distribution unlabeled samples in the fine-tuning phase. Specifically, the in-distribution samples are generated from the corresponding datasets using data augmentation techniques. The out-of-distribution samples of CIFAR10 (resp. CIFAR100) are randomly selected from CIFAR100 (resp. CIFAR10). We use 25000 unlabeled samples and fine-tune the watermarked models for 10 epochs.

\subsection{Effectiveness of Our Attack}
We consider two strategies: the first one is to just use input preprocessing to remove watermarks. The attacker can choose PST transformation, or PST without imperceptible pattern embedding (denoted as ST). For comparisons, we also implement some state-of-the-art transformations for adversarial example mitigation as baselines: BdR \cite{xu2017feature}, Shield \cite{das2018shield}, PD \cite{prakash2018deflecting}, FD \cite{liu2019feature}, and GB \cite{hendrycks2019benchmarking}. Since they are proposed to remove small perturbations in adversarial examples, we adjust the parameters of these methods to affect samples with large patterns till their accuracy is lower than 70\% (resp. 50\%) on CIFAR10 (resp. CIFAR100).

The second strategy is the end-to-end attack with both PST and fine-tuning. We compare it with the black-box watermark removal attack (FT) in \cite{shafieinejad2019robustness} which only fine-tunes the watermarked models with unlabeled samples. FT takes a set of unlabeled samples as input and obtains their labels using the watermarked model as an oracle. Then, it fine-tunes the watermarked models for 10 epochs with the new sample-label pairs. To be comprehensive, we implement the two solutions with in-distribution data for fine-tuning as well (Proposed-In and FT-In).

Since these input transformation functions introduce high randomization, we run each experiment five times to obtain the statistical results. We calculate the average performance of the attacks on normal samples $P_{D}$ to reflect their usability. We select the minimum accuracy of the attacks on watermark samples $P_W$ to quantify their effectiveness, which is efficient to reflect the threat an attack can bring. The experimental results lead to the same conclusion with the average accuracy.

The experimental results are illustrated in Table \ref{tab:effectiveness}. For the first strategy, we observe that perturbation and OOD watermarks are vulnerable to all kinds of transformations. The adversarial example transformations (except FD) and PST (also ST) can easily remove these two types of watermarks on both CIFAR10 and CIFAR100. However, adversarial example transformations are not helpful in removing pattern-based watermarks. Although GB can remove Content watermarks on CIFAR10 and CIFAR100, all adversarial example transformations can not affect Patch watermarks even $P_{D}$ is extremely low. In contrast, PST can significantly reduce the accuracy of the watermarked model on all types of watermarks without heavily affecting $P_{D}$.  ST is slightly worse than PST, but is still much better than the other transformations.

For the second strategy, we observe that although FT-In and FT-Out have almost the same performance on normal samples, they cannot remove watermarks, which was confirmed by the results in existing watermarking papers \cite{adi2018turning,zhang2018protecting}. Our attack (Proposed-In and Proposed-Out) can remove all watermarks and preserve acceptable $P_{D}$ values. Besides, compared with \AlgName, the proposed fine-tuning process can not only improve the performance of the models on normal samples, but also further cleanse the watermarks.

\begin{table}[!t]\centering
    \resizebox{0.95\columnwidth}{!}{
    \begin{tabular}{llrrrrrrrrr}\toprule
        \multirow{2}{*}{Datasets} &\multirow{2}{*}{Attack} &\multicolumn{2}{c}{Perturbation} &\multicolumn{2}{c}{Patch} &\multicolumn{2}{c}{Content} &\multicolumn{2}{c}{OOD} \\
        \cmidrule(lr){3-10}
        & & $P_D$ &$P_W$ &$P_D$ &$P_W$ &$P_D$ &$P_W$ &$P_D$ &$P_W$ \\
        \midrule
        \multirow{12}{*}{CIFAR10} &None &0.92 &1.00 &0.93 &1.00 &0.94 &1.00 &0.92 &1.00 \\
        &BdR &0.45 &\textbf{0.36} &0.48 &1.00 &0.48 &0.89 &0.47 &\textbf{0.11} \\
        &Shield &0.67 &\textbf{0.34} &0.63 &1.00 &0.61 &0.97 &0.66 &\textbf{0.11} \\
        &PD &0.51 &\textbf{0.47} &0.56 &0.94 &0.54 &0.80 &0.53 &\textbf{0.11} \\
        &FD &0.57 &\textbf{0.61} &0.59 &1.00 &0.59 &0.53 &0.55 &\textbf{0.15} \\
        &GB &0.54 &\textbf{0.46} &0.66 &0.98 &0.62 &\textbf{0.06} &0.66 &\textbf{0.15} \\
        &ST &0.83 &\textbf{0.46} &0.86 &\textbf{0.11} &0.84 &\textbf{0.11} &0.83 &\textbf{0.06} \\
        &PST &0.82 &\textbf{0.43} &0.85 &\textbf{0.09} &0.84 &\textbf{0.11} &0.82 &\textbf{0.06} \\
        \cmidrule(lr){2-10}
        &FT-In &\textbf{0.91} &1.00 &\textbf{0.92} &0.94 &\textbf{0.92} &1.00 &\textbf{0.91} &0.70 \\
        &FT-Out &\textbf{0.89} &0.97 &\textbf{0.90} &0.59 &\textbf{0.91} &0.79 &\textbf{0.89} &0.71 \\
        &Proposed\_In &\textbf{0.89} &\textbf{0.34} &\textbf{0.91} &\textbf{0.01} &\textbf{0.91} &\textbf{0.05} &\textbf{0.89} &\textbf{0.04} \\
        &Proposed\_Out &\textbf{0.87} &\textbf{0.47} &\textbf{0.89} &\textbf{0.03} &\textbf{0.89} &\textbf{0.06} &\textbf{0.87} &\textbf{0.08} \\
        \midrule
        \multirow{12}{*}{CIFAR100} &None &0.74 &1.00 &0.74 &1.00 &0.75 &1.00 &0.73 &1.00 \\
        &BdR &0.16 &\textbf{0.05} &0.14 &0.79 &0.16 &0.56 &0.15 &\textbf{0.00} \\
        &Shield &0.30 &\textbf{0.15} &0.30 &0.97 &0.29 &0.61 &0.32 &\textbf{0.02} \\
        &PD &0.25 &\textbf{0.32} &0.24 &0.44 &0.25 &0.41 &0.23 &\textbf{0.01} \\
        &FD &0.33 &\textbf{0.20} &0.33 &0.27 &0.33 &0.31 &0.32 &\textbf{0.03} \\
        &GB &0.39 &\textbf{0.13} &0.42 &0.91 &0.41 &\textbf{0.01} &0.43 &\textbf{0.02} \\
        &ST &0.60 &\textbf{0.20} &0.61 &0.20 &0.61 &\textbf{0.08} &0.61 &\textbf{0.01} \\
        &PST &\textbf{0.69} &\textbf{0.20} &\textbf{0.69} &0.20 &\textbf{0.69} &\textbf{0.07} &0.68 &\textbf{0.00} \\
        \cmidrule(lr){2-10}
        &FT-In &\textbf{0.71} &0.93 &\textbf{0.71} &1.00 &\textbf{0.70} &0.95 &\textbf{0.71} &0.85 \\
        &FT-Out &0.66 &0.73 &0.67 &0.74 &0.67 &0.88 &0.66 &0.75 \\
        &Proposed\_In &\textbf{0.73} &\textbf{0.18} &\textbf{0.74} &\textbf{0.03} &\textbf{0.73} &\textbf{0.05} &\textbf{0.73} &\textbf{0.00} \\
        &Proposed\_Out &\textbf{0.69} &\textbf{0.16} &\textbf{0.70} &\textbf{0.05} &\textbf{0.69} &\textbf{0.07} &\textbf{0.69} &\textbf{0.00} \\
        \bottomrule
        \end{tabular}}
        \caption{Experimental results of watermark removal attacks on normal samples and watermark samples.}\label{tab:effectiveness}
    \end{table}

\subsection{Impact of Parameters}
\begin{figure}[t]
    \centering
    \includegraphics[width=0.45\columnwidth]{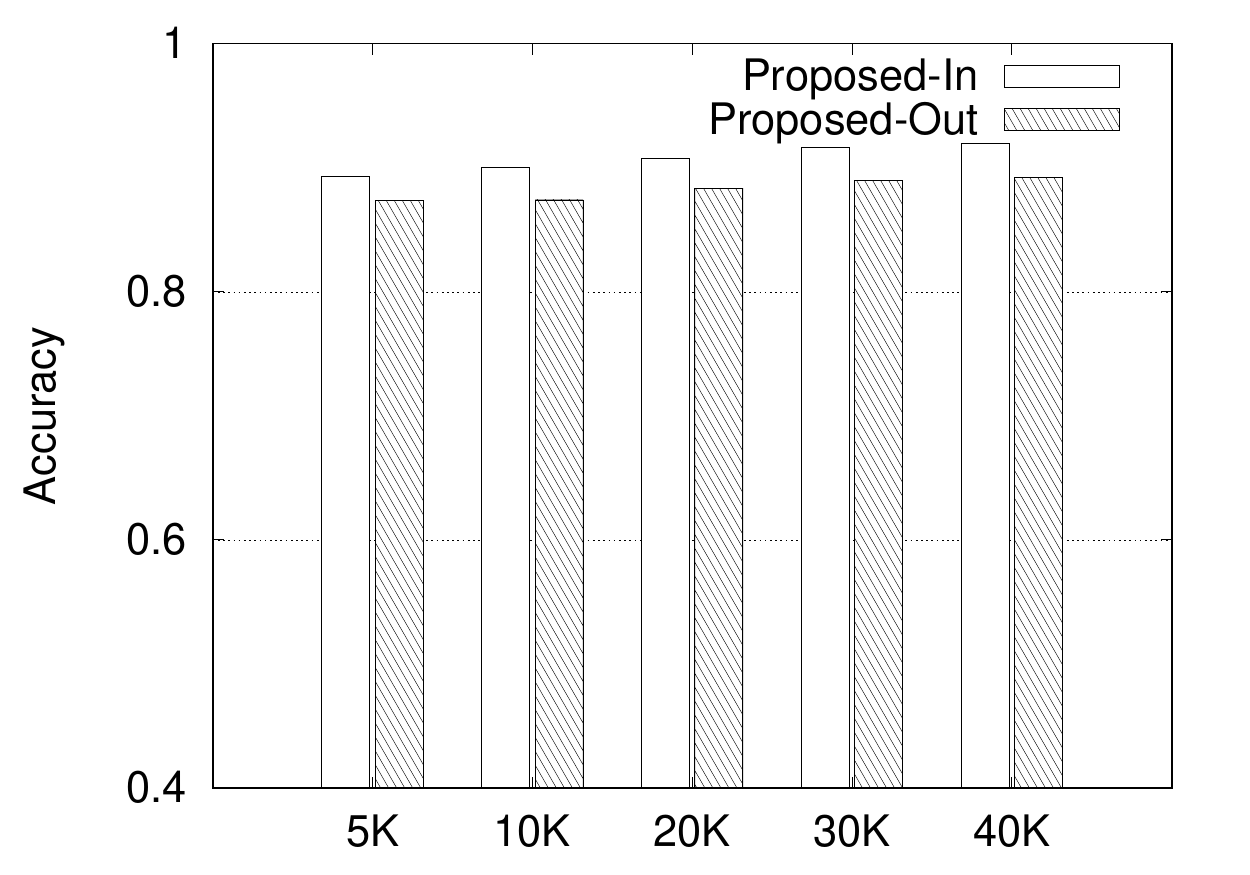}
    \includegraphics[width=0.45\columnwidth]{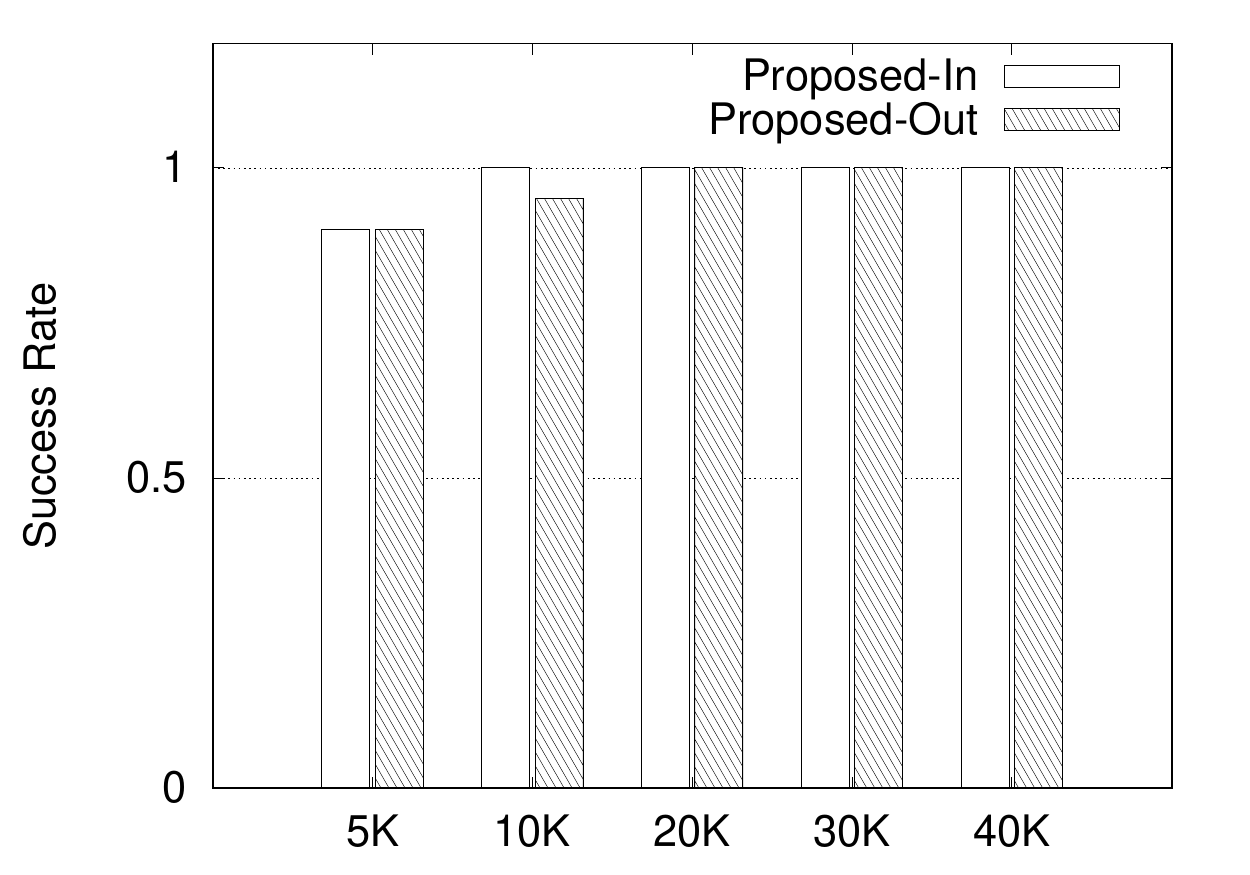}
    \caption{Performance of our attack with different numbers of unlabeled in-distribution and out-of-distribution samples.}
    \label{fig:sample}
\end{figure}

\bheading{Varying the number of unlabeled samples.} We evaluate the performance of our attack with different numbers of unlabeled samples on CIFAR10. Figure \ref{fig:sample} shows the performance of our attack on Patch watermarks. We observe that there exists a trade-off: more fine-tuning samples can increase the performance, which is harder to obtain. If the adversary has enough unlabeled samples, he can obtain a high-quality model with the same performance as the watermarked one. Beside, the success rates of our attack are preserved and even slightly increases with more unlabeled samples, which indicates its severe threat to existing watermarking schemes.

\begin{figure}[t]
    \centering
    \includegraphics[width=0.45\columnwidth]{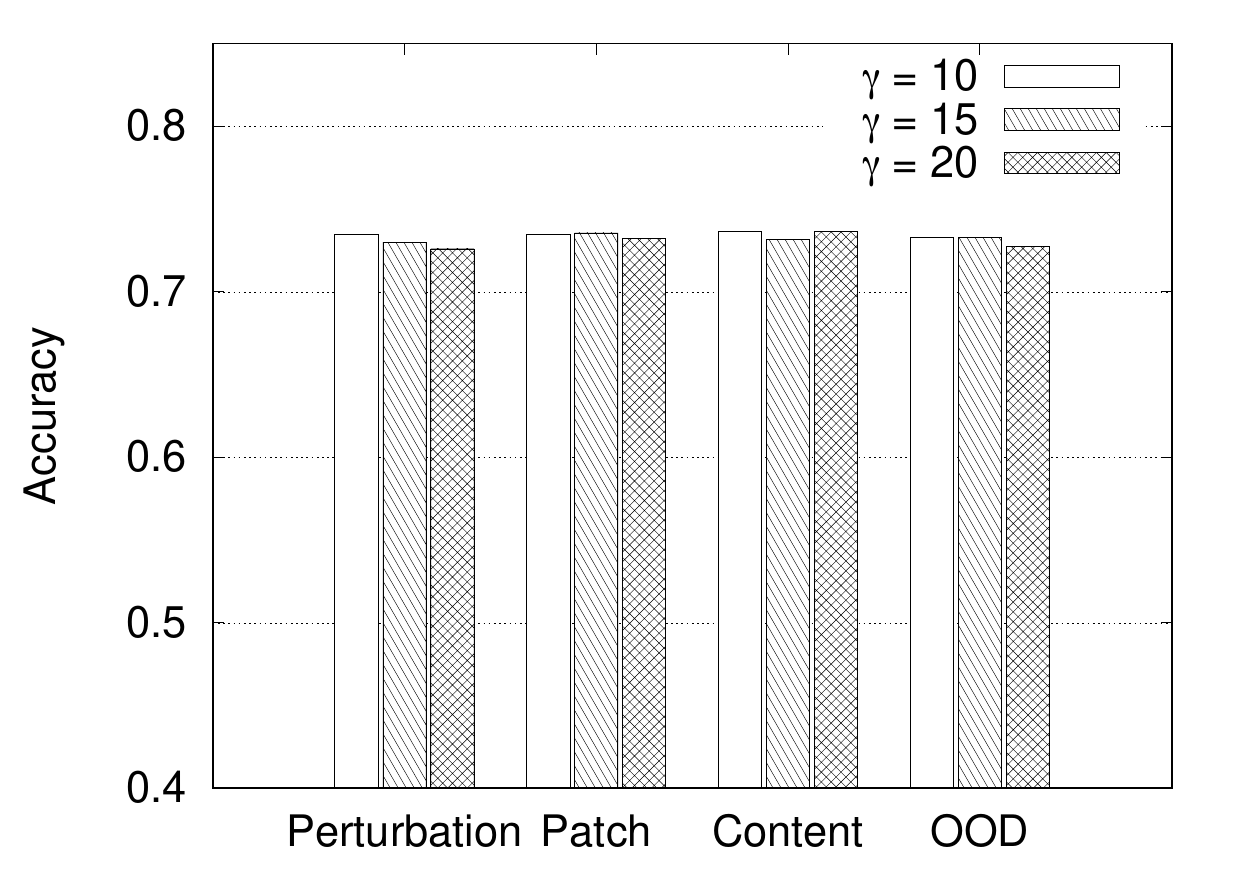}
    \includegraphics[width=0.45\columnwidth]{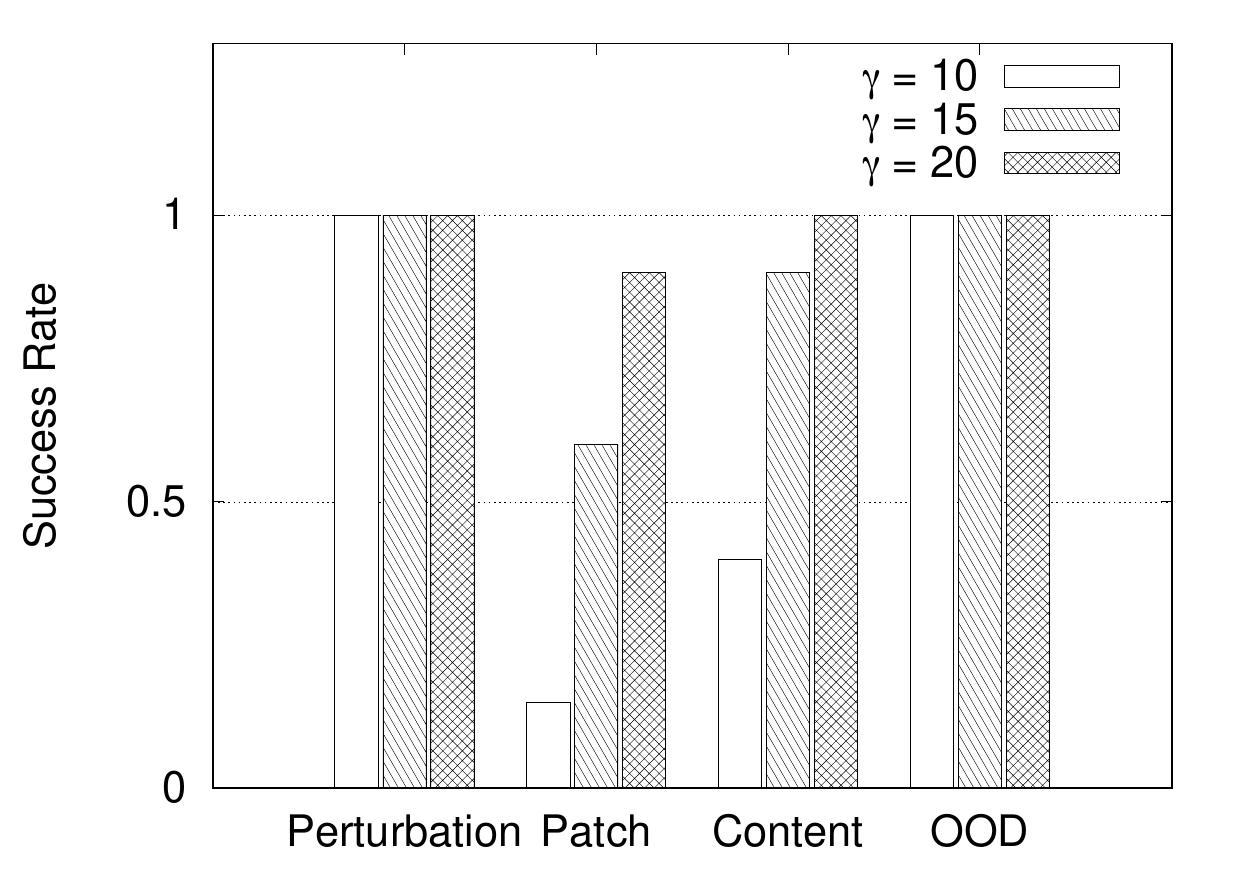}
    \caption{Performance of Proposed-Out with different $\gamma$ values on CIFAR100.}
    \label{fig:gamma}
    \vspace{-10pt}
\end{figure}

\bheading{Varying $\gamma$.} Figure \ref{fig:watermarks} visually illustrates the transformed output of normal and watermark samples with $\gamma = 10$ (second row) and $\gamma = 20$ (third row). Two observations are made from this figure. First, the distortion caused by \AlgName increases as the displacement bound $\gamma$ increases. Second, the samples are randomly preprocessed, which indicates that our attack can be adopted for various watermarking schemes with different settings. We also measure the performance of our attack with different $\gamma$ values on CIFAR100. The experimental results are presented in Figure \ref{fig:gamma}. We observe that due to the effect of the distortion, the model accuracy decreases (except Content) and the success rate increase on the four types of watermarks with an increased $\gamma$.

\section{Watermark Augmentation for Robustness Enhancement}
As discussed in previous sections, inference preprocessing with model fine-tuning can effectively remove state-of-the-art watermarks from the protected model. So we aim to seek for new solutions to enhance the robustness of existing watermarking schemes. Inspired by the data augmentation techniques that can improve the generalization ability of DNN models, we propose \emph{watermark augmentation} to strengthen the watermarks during the embedding process.

\bheading{Methodology}.
A variety of transformations \cite{iccv2019autoaugment} have been designed for data augmentation. Given a specific watermarking scheme with a set of watermarks, we apply those transformations over each watermark sample to augment the watermark set. Then we follow the \textbf{\textit{Mark}} procedure to embed the original and transformed samples into the model. With this watermark augmentation method, the protected model is expected to recognize the watermark samples even the model and samples are perturbed by the adversary.


\bheading{Evaluations}.
We leverage three augmentation transformations: \texttt{ShearX}, \texttt{Rotate}, and \texttt{PST}. Each function has different strength values to transform images. At each iteration, we randomly sample an augmentation strength value and apply the selected transformation to the watermark samples, before embedding them to the model.

We measure the effectiveness of the watermark augmentation strategy against our \AlgName on CIFAR10 and CIFAR100.
Table \ref{tab:defense} illustrates the experimental results of the three augmentation defenses. We observe that watermark augmentation can significantly improve the robustness of the watermarks (expect Rotate on CIFAR10) compared to the results without any defense in Table \ref{tab:effectiveness}. Besides, watermark augmentation with \AlgName is more effective than other transformations.

\begin{table}[t]\centering
    \resizebox{0.95\columnwidth}{!}{
    \begin{tabular}{llrrrrrrrrr}\toprule
        \multirow{2}{*}{Datasets} &\multirow{2}{*}{Augmentation} &\multicolumn{2}{c}{Perturbation} &\multicolumn{2}{c}{Patch} &\multicolumn{2}{c}{Content} &\multicolumn{2}{c}{OOD} \\
        \cmidrule(lr){3-10}
        & & $P_D$ &$P_W$ &$P_D$ &$P_W$ &$P_D$ &$P_W$ &$P_D$ &$P_W$ \\
        \midrule
        \multirow{3}{*}{CIFAR10} &ShearX &0.84 &0.90 &0.86 &0.84 &0.84 &0.58 &0.85 &0.81 \\
        &Rotate &0.84 &0.85 &0.84 &0.23 &0.84 & \underline{0.15} &0.85 &0.73 \\
        &PST &0.80 &1.00 &0.79 &1.00 &0.81 &1.00 &0.84 & 0.99 \\
        \midrule
        \multirow{3}{*}{CIFAR100} &ShearX &0.68 &0.63 &0.70 &0.70 &0.68 &0.43 &0.69 &0.77 \\
        &Rotate &0.68 &0.64 &0.68 &0.20 &0.69 &0.21 &0.68 &0.62 \\
        &PST &0.69 &0.99 &0.68 &0.94 &0.66 &0.96 &0.70 &1.00 \\
        \bottomrule
        \end{tabular}}
        \caption{Experimental results of watermark augmentation defenses against our attack on normal and watermark samples.}\label{tab:defense}
    \end{table}

\section{Conclusion and Future Work}
\label{sec:conclusion}
In this paper, we propose an efficient and practical attack to blindly remove watermarks in DL models. We introduce a novel preprocessing function \AlgName, to transform samples for model fine-tuning and preprocess inference samples. This function can invalidate the effects of watermark perturbations and patterns, without compromising the model usability on normal samples. Extensive experiments demonstrate that the proposed attack technique can remove various types of state-of-the-art watermarks without any prior knowledge about the watermarking schemes and labeled training samples. This indicates existing watermarking schemes still have high severity vulnerabilities. We also propose a new defense strategy to enhance the robustness of existing watermark mechanisms.

For the attack, we empirically and comprehensively validate our technique, following all existing relevant works. Unfortunately, we cannot formally prove the effectiveness of our solutions against all watermarking schemes. Formal and theoretical analysis about watermarks and attacks is an open and challenging problem. To our best knowledge, currently there are no works or attempts to address this problem yet. We will consider this as an important direction for future work. We will also evaluate our attack as a defense against adversarial examples and backdoors.

For the defense, evaluations show the model with augmented watermarks is more robust against simple attacks. The possibility of adaptive attacks targeting this defense is unknown and worth exploration. We hope our work can heat up the arms race between the model owner and plagiarist to inspire the designs of more advanced watermark schemes and attacks in the future.

\section*{Acknowledgments}
This research was supported by the National Research Foundation, Singapore, under its National Cybersecurity R$\&$D Programme (NCR Award Number NRF2018NCR-NCR009-0001 and NRF2018NCR-NCR005-0001), Singapore Ministry of Education AcRF Tier 1 RS02/19 and 2018-T1-002-069, the Singapore National Research Foundation under NCR Award Number NRF2018NCR-NSOE003-0001, NRF Investigatorship NRFI06-2020-0022, the National Natural Science Foundation of China under Grants 62072062 and U20A20176, and the Natural Science Foundation of Chongqing, China, under Grant cstc2019jcyjjqX0026.
\bibliographystyle{named}
\bibliography{body/ref}

\end{document}